\documentclass[aps,prb,twocolumn,groupedaddress]{revtex4}
\usepackage{amsmath,amssymb,epsfig,color}
\usepackage{float}

\begin{document}

\title{Fermi surface nesting and the origin of Charge Density Waves in metals}
\author{M.D. Johannes, I.I. Mazin}
\affiliation{Code 6393, Naval Research Laboratory, Washington, D.C. 20375}

\begin{abstract}
The concept of a CDW induced by Fermi-surface nesting originated from the
Peierls idea of electronic instabilities in purely 1D metals and is now
often applied to charge ordering in real low-dimensional materials. The idea
is that if Fermi surface contours coincide when shifted along the observed
CDW wave vector, then the CDW is considered to be nesting-derived. We show
that in most cases this procedure has no predictive power, since Fermi
surfaces either do not nest at the right wave vector, or nest more strongly
at the wrong vector. We argue that only a tiny fraction, if any, of the
observed charge ordering phase transitions are true analogues of the Peierls
instability because electronic instabilities are easily destroyed by even
small deviations from perfect nesting conditions. Using prototypical CDW
materials NbSe$_2$, TaSe$_2$, and CeTe$_3$, we show that such conditions are
hardly ever fulfilled, and that the CDW phases are actually structural phase
transitions, driven by the concerted action of electronic and ionic
subsystems, \textit{i.e.,} \textbf{q}-dependent electron-phonon coupling
plays an indispensable part. We also show mathematically that the original
Peierls construction is so fragile as to be unlikely to apply to real
materials. We argue that no meaningful distinction between a CDW and an
incommensurate lattice transition exists.
\end{abstract}

\maketitle

One common misconception in modern solid state physics is that Fermi surface
(FS) nesting is always, or nearly always responsible for a charge density wave (CDW).
While many materials experience a structural or magnetic transition with a
wave vector that is incommensurate or badly commensurate with the
high-symmetry phase, and while in some cases a visual inspection of the FS
seems to reveal nesting parts with roughly the same wave vector, a
quantitative search aimed at finding a nesting-driven instability at the
experimental CDW vector on the level of one-electron energies practically
always fails \cite{NbTe2, Hidden,bronze,Mattheis,aebi}. This failure is a
symptom of a larger misconception about CDW's, specifically, that they are
the result of a purely \textit{electronic} instability along the lines of
the Peierls instability in one dimension (1D), and an even bigger
misconception that the structure of the electronic susceptibility in the
reciprocal space can be revealed by inspecting the Fermi surface alone,
without analyzing the high-energy electronic excitations.

In the Peierls picture, lattice distortion is a secondary effect that arises in response to an electronically driven charge 
redistribution that would occur regardless of whether or not the ions subsequently shift from their high symmetry positions. 
In real materials, the electronic and ionic instabilities always occur simultaneously. Computational attempts to stabilize a 
CDW without allowing the ions to move have failed in all cases we are aware of, most particularly for protototypical CDW 
metal NbSe$_2$. We will show that the concurrence of the two transitions is not a coincidence but arises from the fact that 
CDW formation relies on the lattice distortion as an essential element and not the reverse. The necessity of strong 
\textbf{q}-dependent electron-phonon coupling indicates that Fermi surface nesting, a purely electronic effect, may help CDW 
formation, but cannot be the only driving force behind the CDW phenomenon \cite{chan}.

In the literature the term CDW is used in two different senses. In some
cases, a structural transition with an incommensurate or long period is
termed a CDW, regardless of its origin, while in other cases, the words CDW
are reserved for Peierls-like instabilities that occur due to a divergency
in the real part of the electronic susceptibility, so that the electronic
subsystem would be unstable \textit{per se}, even if the ions were clamped
at their high symmetry positions. We choose the latter definition for two
reasons: first, because many accepted CDW materials actually exhibit a
commensurate CDW phase \cite{dical}, and second, because it allows for a
distinction between a general incommensurate lattice transition (ILT) and a
CDW. Moreover, the archetypal CDW, the 1D Peierls transition in a
half-filled band, is commensurate with just a doubled unit cell. To fit the
definition of a Peierls system (and therefore for a CDW, for the purposes of
this paper) a system must satisfy several requirements: (a) there must be
substantial nesting of the FS. Note that a quantitative measure of the FS
nesting, sometimes called the "nesting function" is nothing but the
low-frequency limit of the imaginary part of the bare electronic
susceptibility, $\chi _{0}^{\prime \prime }(\mathbf{q)}$, in the constant
matrix element approximation \cite{nesting}. This must, correspondingly,
peak at the CDW wave vector. (b) the nesting-derived peak must carry over
into the real part of the susceptibility, $\chi _{0}^{\prime }(\mathbf{q)}$,
at the same wave vector, because it is $\chi ^{\prime }(\mathbf{q)}$ that
defines the stability of the electronic subsystem, (c) the peak in $\chi
_{0}^{\prime }(\mathbf{q)}$ must translate into a divergence in the
full electronic susceptibility to cause the electronic subsystem to be
unstable even without any ionic shifts, and (d) \textit{all} phonons must
soften at the CDW wave vector, not only the one corresponding to the mode
that eventually gives rise to the observed CDW (except maybe a few that
cannot couple to this electronic instability for a particular symmetry
reason). With respect to the final point, we are not aware of any material
in which such softening for multiple modes has actually been observed, although in theory it is
unavoidable\cite{multimode}. In fact, we intend to show that this definition
as a whole is not fulfilled in real systems and therefore, aside from
convention, there is nothing to distinguish a CDW from a structural phase
transition.

The issue is even more confounded by the fact that the words \textquotedblleft
nesting\textquotedblright\ and \textquotedblleft Peierls
transition\textquotedblright\ are also interpreted differently by different
researchers.  We use the first term in its literal sense: Fermi surface parts
are nested if, when shifted rigidly, they coincide with other Fermi surface
parts. Some authors, appreciating the fact that this is not sufficient for a
peak in  $\chi ^{\prime }(\mathbf{q)}$, distinguish between
\textquotedblleft real nesting\textquotedblright\ that exists not only at
zero frequency, but also within some finite range of transition energies (as
we discuss later in connection with CeTe$_{3},$ this imposes an additional
constraint on the Fermi velocities), and \textquotedblleft false
nesting\textquotedblright\ that has infavorable Fermi velocities. As for the second term,
there is a tendency to use Peierls transition as a synonym for
dimerization ($e.g.,$ dimerization in VO$_{2}$ is often called a Peierls
transition \cite{silke2}). 
Again, we adhere to a more strict definition that reserves this nomenclature for a
transition driven by a lowering of one-electron band energy caused by the opening of a gap, as in the original Peierls model. Note that 
the doubling of the unit
cell may occur via dimerization, but can also easily occur via zigzag-ing
of the atomic chain or via an even more complicated pattern.

In Section I, we consider the classical 1D Peierls transition, for which the conditions listed above are fully satisfied and 
in which FS nesting is indeed expected to give rise to CDW formation (the results for a nested 2D system, such as the nearest 
neighbor one-band TB model, are qualitatively the same). We will demonstrate how fragile this construction is even in 1D, 
since a CDW of fully electronic origin is exponentially weak. Furthermore, we will show how rather small deviations (on the 
order of what is expected in real materials) from the perfect model are sufficiently strong to suppress FS nesting-driven CDW 
formation. In Section II, we perform density functional calculations of several real materials that are commonly considered 
to be CDW systems and show that, even in these canonical systems, electron-phonon coupling and not nesting is at the heart of 
the CDW phenomenon. For clear understanding in both following sections, it is necessary to point out that an electronic CDW 
instability is not induced by a divergence in the imaginary part of susceptibility, $\chi ^{\prime \prime }(\mathbf{q}),$
which is the function that reflects the FS topology and can be easily measured experimentally by neutron scattering.  
Instead, it is the real part, $\chi ^{\prime }(\mathbf{q})$, which must diverge in order to trigger an electronic CDW.  
Unfortunately, $\chi ^{\prime }(\mathbf{q})$, is difficult to map experimentally (see, however, Ref \onlinecite{borisenko}). 
We write the two parts, in the constant matrix element approximation as:

\begin{eqnarray} \chi ^{\prime }(\mathbf{{q})} 
&\mathbf{=\sum_{k}}&\frac{f(\varepsilon _{ \mathbf{k}})-f(\varepsilon _{\mathbf{k+q}})}{\varepsilon _{k}-\varepsilon _{ 
\mathbf{k+q}}} \label{compsusc} \\ \lim_{\omega \rightarrow 0}\chi ^{\prime \prime }(\mathbf{q},\omega )/\omega 
&=&\sum_{k}\delta (\varepsilon _{k}-\varepsilon _{F})\delta (\varepsilon _{k+q}-\varepsilon _{F}) \end{eqnarray}

Since $\chi ^{\prime \prime }(\mathbf{q,\omega \rightarrow 0})$ is easier to
calculate, it is often presented in first-principle studies as a
quantitative test of the FS nesting (which it is) and/or as a gauge of a
tendency to CDW formation (which it is not) \cite{clerc,koitzsch,landa,kunes,kasinathan,kolesnychenko}.

\section{A model Peierls system}

In the 1D Peierls system, we have a parabolic band of noninteracting particles in a periodic 
external potential that disperses as $E=k^{2}$ (in Ry units) and has a particular Fermi vector 
$\mu =k_{F}^{2}$, where $\mu $ is the chemical potential. One assumes that the band is half 
filled so that the first reciprocal lattice vector $G=2\pi /a=4k_{F}.$ Peierls was the first to 
point out that such a system is formally unstable against any doubling of the unit cell, 
because it creates an additional potential with a non-zero component at $q=2k_{F},$ $V(q)\neq 
0$ and opens a gap. He emphasized that a gain in the one-electron energy depends on the 
amplitude of lattice distortion logarithmically (as $u^{2}\log u),$ while the elastic energy is 
normally quadratic, and therefore the ground state would always correspond to a nonzero 
distortion. It appears that actual numerical calculations do not necessarily find a distortion 
\cite{yossi}. Ashkenazi {\it et al} \cite{yossi} argue that when the electronic susceptibility 
is nonanalytic, the elastic energy may be as well and therefore it is not guaranteed that for 
an infinitesimal distortion, the one-electron energy will be larger.  This same point can be 
argued as an inability to cleanly partition the total energy into the one-electron energy and 
the elastic energy in a real system, where the particles in question are interacting electrons. 
Instead, the full expression for the total energy (in density functional theory, for instance) 
must be analyzed. Following Ref. \onlinecite{yossi}, we can write, using implicit matrix 
notation (in real or reciprocal space) the change of energy arising from the nuclear 
displacement in terms of the change of the potential of the nuclei, $\delta V_{ext}$, and the 
induced density change, $\delta n$, as: \begin{eqnarray} \delta E_{tot}
&=&-(1/2)\delta n\chi ^{-1}\delta n \\ \delta n &=&\chi \delta V_{ext} \\ \chi &=&\chi _{0}/\epsilon \\ \epsilon 
&=&1-v_{i}\chi _{0}=(1+v_{i}\chi )^{-1} \\ \delta E_{tot} &=&-(1/2)\delta V_{ext}\chi \delta V_{ext} \\ &=&-(1/2)\delta 
V_{ext}\chi _{0}(1-v_{i}\chi _{0})^{-1}\delta V_{ext}, \end{eqnarray} where $v_{i}$ is the total DFT interaction, including the Coulomb and exchange-correlation kernel. Note 
that we use a sign convention such that in a stable system $\chi _{0}<0.$ Neglecting the matrix character of these equations, we immediately observe that a divergence in 
$\chi _{0},$ as found by Peierls, is by itself insufficient to cause an instability, since the total susceptibility is bounded by $ -v_{i}$ and does not diverge. This is not 
necessarily true if the matrix character (or local fields) is taken into account and such an inclusion results in a complicated formula for the phonon frequencies, known as 
the Pick-Cohen-Martin formula as discussed in Ref. \onlinecite{Nb_us}.  However, without these considerations, this result demonstrates that there is no direct relation 
between the Peierls instability in a system of noninteracting particles and CDWs in real systems.  The same point can be made using a linear response approach \cite{yeats}: 
an interacting half-filled electronic system is stable against infinitesimal perturbation. Only a finite distortion, and only if e-ph coupling is larger than a critical 
value, can be stable.

This is already a very serious reservation, but nonetheless it is instructive to step back and investigate the 
\textquotedblleft 
classical\textquotedblright\ Peierls instability in a noninteracting system. There is no question that the susceptibility of this system 
is logarithmically divergent, but there are interesting, and important questions left to ask.  First, is this divergency robust with 
respect to small deviations from a \textquotedblleft perfect nesting\textquotedblright , as is always the case in real materials? Second, 
where is the energy gain associated with this instability collected? That is, can electrons away from the Fermi surface be effectively 
neglected or must the effects of lower filled states be taken into account?

The standard expression for the real part of susceptibility reads 
\begin{equation}
\chi ^{\prime }(q)=\frac{1}{q}\ln \left\vert \frac{q-2k_{F}}{q+2k_{F}}
\right\vert .  \label{ideal}
\end{equation}
This expression is normalized to $1/k_{F}$ at $q\rightarrow 0$ (the overall
scale is not important at the moment) and has a very weak, logarithmic
divergency at $q=\pm 2k_{F}.$ To illustrate just how weak it is, we assume a
relaxation rate $\gamma ,$ corresponding to the Drude relaxation rate in
optics, and recalculate $\chi ^{\prime }(q)$. The new result reads: 
\begin{equation}
\chi ^{\prime }(q)=\frac{1}{2q}\ln \left\vert \frac{\gamma
^{2}+q^{2}(q-2k_{F})^2}{\gamma ^{2}+q^{2}(q+2k_{F})^2}\right\vert .
\label{gamma}
\end{equation}
The divergency has been reduced to merely an enhancement of $\chi ^{\prime
}(\pm 2k_{F})$ over $\chi ^{\prime }(0)$ by a factor of $\ln (1+\frac{64\mu
^{2}}{\gamma ^{2}})/4.$ For typical $\gamma $ of the order of 0.1-0.2 eV
this enhancement is by a factor of 2-2.5. One can also add that at any
finite temperature, even without relaxation, $\chi ^{\prime }(\pm
2k_{F})/\chi ^{\prime }(0)=\ln ( \frac{2k_{F}^{2}-kT}{kT})/2\approx \ln (
\frac{2\mu }{kT})/2.$ For typical Fermi energies and T=10K (most observed
ILTs occur at higher temperatures) the enhancement is again only a factor of
the order of four.

Thus, carrier scattering and Fermi function broadening are suffucient to
reduce the nesting-induced divergency to a minor structure in $\chi ^{\prime
}(q).$ But an even more severe effect is caused by geometrical deviations of
the Fermi surface from perfect nesting. A common procedure in the search for
nesting vectors in a particular fermiology is to copy a quasi-2D Fermi
surface cut onto transparent paper and slide it until some piece of the
displaced Fermi surface visually coincides with another piece of the
original plot. It is instructive to give a quantitative gauge of what
constitutes a "good nesting" vs. a "bad nesting". Assuming that the
\textquotedblleft nested\textquotedblright\ parts really nest only up to
some $\delta k$ in the reciprocal space, we observe, by averaging Eq.\ref{ideal}, that 
\begin{equation}
\begin{split}  \label{2D}
&\chi ^{\prime }(q) =\frac{1}{2\delta k}\ln \left\vert \frac{
q^2-(2k_F-\delta k)^{2}}{q^2 +(2k_F -\delta k)^{2}}\right\vert + \\
&\frac{k_{F}}{\delta k q} \ln \left\vert \frac{(q-\delta k)^{2}-4 k_F^{2}}{
(q+ \delta k)^{2}-4 k_F^{2}}\right\vert + \frac{1}{2q}\ln \left\vert \frac{
(q- 2k_F)^{2}- \delta k^{2}}{(q+2k_F)^{2}- \delta k^{2}}\right\vert \\
\intertext{which gives:}
&\chi ^{\prime }(2k_{F}) =\frac{1}{\delta k}\ln \left\vert \frac{4k_{F} -
\delta k}{4k_F +\delta k}\right\vert +\frac{1}{4k_F}\ln \left\vert \frac{
\delta k^{2}}{\delta k^{2}-16k_F^2}\right\vert
\end{split}
\end{equation}

\begin{figure}[]
\includegraphics[width = 0.95\linewidth]{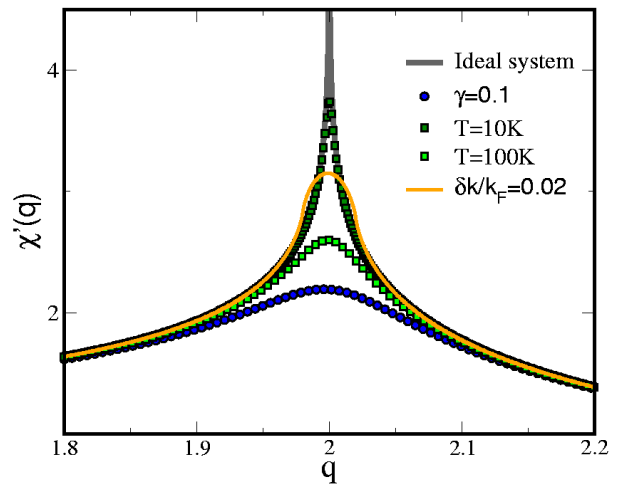}
\caption{(color online) A comparison of $\protect\chi ^{\prime} (q)$ under
ideal 1D conditions with perfect nesting at T=0 to $\protect\chi ^{\prime}
(q)$ under various non-ideal conditions. Even moderate deviations from the
ideal (such as those found in real materials) rapidly reduce the divergence
to a relatively weak enhancement.}
\label{decrease}
\end{figure}

In the small $\delta k$ limit, $\chi ^{\prime }(2k_{F})/\chi ^{\prime }(0)\approx 1/2[(1+\ln (4k_{F}/\delta k)].$ In other 
words, if 2D Fermi lines coincide within 5\% of the Fermi vector, the corresponding enhancement of the susceptibility is 
about a factor of three.  It is also important to remember that real materials are \textit{quasi}-2D, not 2D. Any dispersion in the third direction of the order of $\delta k$
brings us again to Eq.\ref{2D}. In a one-band case one can estimate the
ratio $k_{F}/\delta k$ as $\omega _{p\perp }^{2}/\omega _{p||}^{2}\approx
\rho _{\perp }/\rho _{||}.$ This shows that anything with a calculated
transport anisotropy of less than one order of magnitude is 3D from a
\textquotedblleft nesting point of view\textquotedblright. Fig.\ref{decrease}
illustrates the effect of various deviations from the perfect Peierls
picture on the divergence of $\chi ^{\prime }(q) $.

To further emphasize the non-role that FS nesting plays, let us investigate
a system in which an ILT with $Q=2k_{F}$ has actually occurred, and as a
result a gap equal to $2V$ has opened, with the corresponding energy gain in
the one-electron energy. The common wisdom is that energy is gained
predominantly near the former Fermi energy, where the lowering of
one-electron states is the largest. But, is this really so? In first order
perturbation theory, the new one-electron spectrum is 
\begin{eqnarray}
E_{k}^{\prime } &=&\frac{E_{k}+E_{k-Q}-\sqrt{[E_{k}-E_{k-Q}]^{2}+4V^{2}}}{2},
\\
\delta E_{k} &=&E_{k}-E_{k}^{\prime }=\frac{\Delta E_{k}+\sqrt{\Delta
E_{k}{}^{2}+4V^{2}}}{2},
\end{eqnarray}
where $\delta E_{k}$ is the energy shift of the state $|k\rangle $ and $
\Delta E_{k}$ is the energy difference between the two states connected by
the nesting vector $Q.$ An inspection of this expression shows that
integrating it over $\Delta E_{k}$ does not diverge at $\Delta E_{k}=0$
(that is, at the Fermi energy), but would diverge at large energies, $\Delta
E_{k}\rightarrow \infty$, if the integration were not limited by the
bandwidth. Substituting $E_{k}=k^{2},$ we get 
\begin{equation}
\delta E_{k}=\sqrt{4k_{F}^{2}(k-k_{F})^{2}+V^{2}}+2k_{F}(k-k_{F}).
\end{equation}
The total energy gain, $E_G = \frac{1}{2k_F} \int_{-k_F}^{k_F} \delta E_k dk 
$ is:

\begin{equation}
\frac{\delta E_{G}}{V^{2}}\approx \frac{1}{16k_{F}^{2}}+\frac{1}{8k_{F}^{2}}
\ln \left( \frac{8k_{F}^{2}}{V}\right) =\frac{1}{16\mu }[1 +2\ln \left( 
\frac{8\mu }{V}\right)] .
\end{equation}
The first term corresponds to integrating over the region $\Delta E_{k}<V,$
and neglecting $\Delta E_{k},$ and the second corresponds to integrating
over the region $\Delta E_{k}>V,$ and neglecting terms smaller than $V^{2}.$
In other words, the first (non-divergent) term comes from states near the
Fermi level, where the gap opens up, and the second term (divergent at large 
$\mu ,$) comes from the rest of the states below the Fermi level and down to
the bottom of the band. Since an actual instability is always a competition
between an electronic (in this picture, one-electron) energy gain and an
elastic energy loss, it is apparently more important to optimize the energy
gain from all occupied states than to open a gap over the largest possible
part of the Fermi surface.

\begin{figure}[tbp]
\includegraphics[width = 0.95\linewidth]{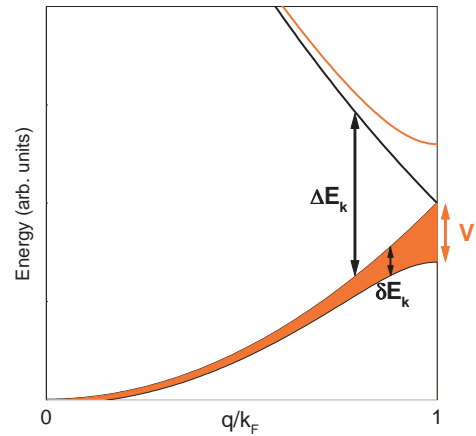}
\caption{(color online) A schematic showing the original energy bands of our
model system in comparison to the bands after a transition has caused a gap
at the Fermi energy. The energy associated with gap formation is shaded
(red) and continues down to the bottom of the band. Representative energy
states connected by the nesting vector are shown ($\Delta E_k$) along with
infinitesimal energy gain ($\protect\delta E_k$) associated with the gap
opening at each \textbf{k}.}
\label{gap}
\end{figure}

Since, as we have just found, all occupied states are important for a ILT,
it is important to keep in mind the multiband character of real solids, and,
correspondingly, the contribution of the interband transitions to $\chi
^{\prime }(\mathbf{q}).$ To get a feeling for the interband effects, let us
consider an insulating system which, instead of the free-electron band
discussed above, has a fully occupied first neighbor tight-binding band, $
E_{1}(k)=-\cos (ka)$ separated from a similar empty band by an energy $
\varepsilon$, $E_{2}(k)=\varepsilon +\cos(ka)$ Let us assume for simplicity
a constant interband dipole matrix element equal to unity. The
susceptibility is given by 
\begin{equation}
\chi _{inter}^{\prime }(q)=\frac{1}{a \sqrt{\varepsilon^2-4\sin ^{2}(qa/2)}}.
\end{equation}
This susceptibility is enhanced at $q=\pi/a$ (at the edge of the BZ)
compared to its minimum value (at $q=0)$ by a factor of $\varepsilon /\sqrt{
\varepsilon ^{2}-4}$. This enhancement can be very large even for bands with
a large relative shift. Using for example $\varepsilon =2.25,$ (giving a gap
of 1/8 of the bandwidth), the enhancement is 2.2, quite comparable to what
one might expect in a realistic nesting scenario. This demonstrates again
that the Fermi surface topology is unlikely to be a factor in CDW formation.
Note that the band structure that we have used is not special in any way, as
opposed to the half-filled band exhibiting a Peierls instability.

\section{First principles calculations of real materials}

In the following sections, we present three examples of real materials,
commonly thought to be canonical examples of nesting-driven CDW systems: NbSe$_2$, TaSe$_2$, and CeTe$_3$. Using first principles density functional
theory calculations, we show that FS nesting is not the driving force behind
the observed CDW in any of the three compounds. We additionally investigate
a chain of Na ions and show that even for this artificially perfect 1D
system, the strong FS nesting fails to produce any appreciable CDW when the
ionic positions are fixed. This is perfectly in line with our contention
that electron-phonon coupling is necessary to instigate the ILT and that
even in the Peierls formulation, the resulting CDW is exponentially small.

For all investigated compounds, the first principles calculations were
performed using the well-known Wien2K package\cite{Wien2k} with the local
density approximation (LDA) to the exchange correlation potential \cite{lda}
. For NbSe$_2$ and TaSe$_2$, it was found that spin-orbit coupling has a
finite effect on the band structure and Fermi surface and therefore was
taken into account. To get a good energy mesh, we calculated eigenvalues at
nearly 15,000 $k$ points in the full BZ. For CeTe$_3$, the partially filled $
f$-states pose a well-known problem for the mean-field LDA methodology by
partially filling each of the $f$ states rather than completely filling some
and emptying all others. We have addressed this shortcoming by using LDA+U
in the fully localized limit scheme \cite{fll}. We applied a U of 4.5 eV and
successfully reproduced the $f^1$ state observed by magnetization
measurements \cite{iyeiri} for the Ce ion, simultaneously removing all $f$
bands from the Fermi energy. None of the quantities we calculate are
expected to be sensitive to the precise value of U and we did not
investigate the effects of changing it. We used a mesh of approximately
30,000 $k$ points in the full BZ to calculate the energy eigenvalues used
for the susceptibilities. For all compounds, we used a temperature smearing
of 2 mRy during numerical integration. For calculation of the chain of Na
atoms we used the Vienna Ab-Initio Simulation (VASP) package with the
projector augmented wave (PAW) basis set \cite{vasp,paw1,paw2}. We used
several different pseudopotentials with varying levels of hardness for the
Na core, all giving identical results to within the accuracy of our
calculations.

\subsection{NbSe$_2$ and TaSe$_2$}

Quasi-two-dimensional NbSe$_2$ belongs to a family of layered
dichalcogenides that undergoes a CDW transition sometimes thought to be
related to FS nesting. A calculation of both the real and imaginary parts of
the one-electron susceptibility \cite{Nb_us} shows this to be unequivocally
not the case. Although FS nesting does exist and does produce a peak in $
\chi ^{\prime \prime }(\mathbf{q})$, it is located at \textbf{$q$} =
(1/3,1/3,0) and not at the observed $\mathbf{q}_{CDW}$ = (1/3,0,0) (See Ref 
\onlinecite{Nb_us} for pictures). On the other hand, a weak peak at $\mathbf{
q}_{CDW}$ appears in $\chi ^{\prime }(\mathbf{q})$. This peak is not strong
enough alone to stimulate the observed CDW transition, but must be assisted
by electron-phonon coupling at the same wave vector. To verify this, we
performed two first principles calculations of NbSe$_2$ in a supercell
corresponding to the observed CDW vector. In one case, we clamp the ions to
their high symmetry positions and in the other we allow them to shift. When
allowed, the ions do indeed shift and an instability at the right wave
vector, (1/3,0,0), is reproduced. In the case where the ions are forced to
be stationary, both ionic and electronic systems remain stable in their high
symmetry states. Even if we prompt the electronic system by artificially
redistributing the charge along the known CDW direction, it relaxes back to
its high symmetry state. Our calculations on this system provide a clear
example of the lack of influence of Fermi surface nesting with respect to
CDWs: a strong FS nesting that produces a sharp peak \textit{fails} to gives
rise to a CDW at the associated wave vector, while a peak wholly unrelated
to FS nesting but at the correct vector appears in the real part of the
susceptibility and, in conjunction with electron-lattice effects (but not
without!) \textit{does} produce a CDW.

\begin{figure}[tbp]
\includegraphics[width = 0.95\linewidth]{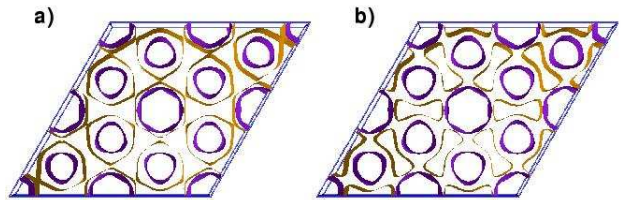}
\caption{(color online) The Fermi surface of TaSe$_2$ shown with a) no shift
of the Fermi level b) a downward shift of the Fermi level (-40 meV). The
topology of the Fermi surface, especially the light (yellow) sheets changes
appreciably with a small change in Fermi energy, and brings the surface into
good agreement with experiment}
\label{TaSe2FS}
\end{figure}

We have also investigated an isostructural material, TaSe$_2$, that has an
observed CDW at approximately the same wave vector as NbSe$_2$. The Fermi
surface of the Ta compound is different from that of NbSe$_2$, as was most
recently observed by Rossnagel \textit{et al} \cite{rossnagel} who also
point out that the measured surface differs from the calculated one \cite
{rossnagel,borisenko,wexler}. This is somewhat surprising because calculated
and observed Fermi surfaces are in quite good agreement for NbSe$_2$. One
reason for the discrepancy between NbSe$_2$ and TaSe$_2$ and between theory
and experiment is spin-orbit coupling, which is strong for the heavy Ta ion.
This coupling has not been included in most previously published TaSe$_2$
band structures and Fermi surfaces \cite{wexler,reshak,barnett}, with the
result that they differ less than they should from NbSe$_2$ band structures.
The spin-orbit interaction non-trivially changes the band dispersion near
the Fermi energy, particularly along $\Gamma-K$ and leads to a different
Fermi surface topology. Scalar relativistic effects are responsible for the
lower Se band in TaSe$_2$ (compared to NbSe$_2$) which removes it from the
Fermi surface entirely. Although we include both scalar relativistic effect
and spin-orbit coupling in our calculations and find qualitative differences
between the Nb and Ta di-selenides, we still find that the TaSe$_2$ Fermi
surface differs somewhat from the ARPES observed surface. However, nearly
perfect agreement can be achieved by a small shift of the Fermi energy
(about 0.04 eV). In Fig. \ref{TaSe2FS}, the Fermi surface both with and
without the small shift in Fermi energy is shown.

\begin{figure}[tbp]
\includegraphics[width = 0.95 \linewidth]{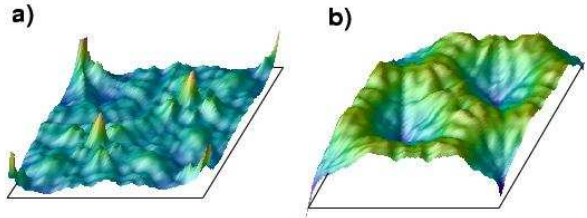}
\caption{(color online) The imaginary (left) and real (right) parts of the
susceptibility for TaSe$_2$. The nesting peaks (imaginary part) do not
correspond to the observed CDW wave vector, while the very weak peaks in the
real part do.}
\label{TASE2}
\end{figure}

The susceptibilities presented here are calculated using the Fermi energy
shift necessary to bring theoretical and experimental Fermi surfaces into
agreement, but in fact this shift has no visible effect whatsoever on either
the real or imaginary parts, despite its dramatic effect on the FS topology.
In Fig.\ref{TASE2}, the real and imaginary parts of the susceptibilities for
TaSe$_2$ are presented. Looking at the susceptibilities in Fig.\ref{TASE2},
one can see some similarity between the peak structures in this compound and
in NbSe$_2$ \cite{Nb_us}, most particularly in the location of the
nesting-driven peak in $\chi ^{\prime \prime }(\mathbf{q})$ vs. the peak in $
\chi ^{\prime }(\mathbf{q})$. Again we find that the peaks are in different
locations, indicating that nesting cannot give rise to a charge instability.

A recent study of these two materials \cite{rossnagel} concluded that both
materials have CDWs driven by strong electron-phonon coupling in the
presence of weak nesting. Our calculations show that, indeed,
electron-phonon coupling is behind the transition, but we can eliminate
Fermi surface nesting from the phenomenon entirely. There is no nesting at
all, not even weak, at the CDW wave vector. The electronic susceptibility
structure, which is indeed favorable to a CDW at the right wave vector, is
due to finite energy electronic transitions and not to a Fermi surface
geometry.

\subsection{CeTe$_3$}

\begin{figure}[tbp]
\includegraphics[width = 0.95\linewidth]{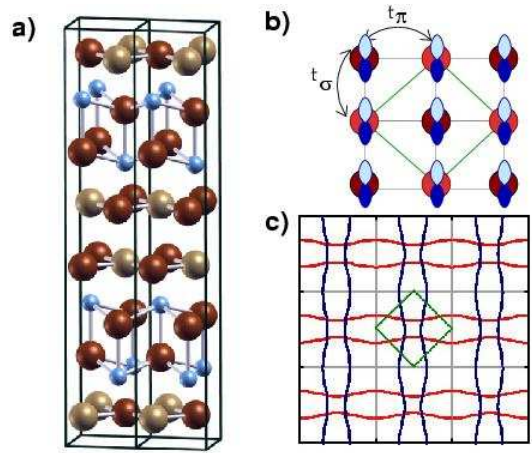}
\caption{(color online) \textit{a}) The structure of CeTe$_3$ showing the Te
(large spheres) planes interposed with Ce (small spheres) and Te staggered
units. One of the two non-equivalent Te ions in the Te-only planes is shaded
lighter to distinguish it. \textit{b}) A schematic of the quasi-1D
tight-binding model employed to illustrate the origin of the Fermi surface. 
\textit{c} The quasi-1D Fermi sheets resulting from the nearest-neighbor
tight binding model. The new BZ corresponding to the larger unit cell in
which the two different Te ions are distinguished is shown as a darker
(blue) diamond.}
\label{cete3struct}
\end{figure}

CeTe$_3$ is another layered material, (see Fig.\ref{cete3struct}) also
belonging to a family of compounds RTe$_3$, R = rare earth, all of which
exhibit CDWs \cite{dimasi,fisher,brouet}. The two-dimensionality is
considerably stronger in this series than in the dichalcogenides discussed
above, producing an easily visible Fermi surface nesting between strongly
two-dimensional Fermi sheets. As in the former series, the strongest nesting
peaks seen in $\chi ^{\prime \prime }(\mathbf{q})$ produced by this nesting
are far away from $\mathbf{q}_{CDW}$ and do not carry over into $\chi
^{\prime }(\mathbf{q})$. However, much weaker peaks associated with a
different nesting do appear in $\chi ^{\prime \prime }(\mathbf{q})$ at the
observed CDW wave vector and are then strongly enhanced by contributions
away from the Fermi energy to eventually produce peaks in $\chi ^{\prime }(
\mathbf{q})$. It is an interesting and instructive exercise to track down
the origin of both kinds of nesting peaks to examine why the strongest ones
are irrelevant and why the weaker peaks are present throughout the energy
spectrum, finally resulting in a peak in $\chi ^{\prime }(\mathbf{q})$. To
begin, we describe how the structure of the CeTe$_3$ very simply gives rise
to its Fermi surface. CeTe$_3$ is composed of two different types of layers:
those containing staggered Ce and Te ions and those containing Te ions only
(see Fig.\ref{cete3struct}). The states near the Fermi energy have
predominantly Te $p$ character and come from the pure Te layers. Assuming a
nearest neighbor only tight binding model of Te $p_x$ and $p_y$ orbitals
(See Fig.\ref{cete3struct}b) in these layers with $t_{\sigma}$ $\approx$ 5$
t_{\pi}$ as in Fig.\ref{cete3struct}b, we produce the crosshatched pattern
of slightly warped one-dimensional Fermi sheets shown in Fig.\ref
{cete3struct}c. This model and the values of the tight-binding parameters
are very similar to those developed in Ref. \onlinecite{Yao}.

\begin{figure}[tbp]
\includegraphics[width = 0.95\linewidth]{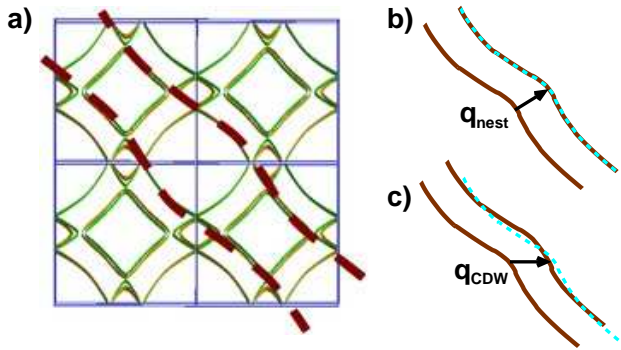}
\caption{(color online) \textit{a)} The full, calculated Fermi surface of
CeTe$_3$ (thin solid lines, green and brown online), with the
one-dimensional bands obtained from a nearest neighbor tight-binding model
overlayed on top (thick dashed crimson lines). The Fermi surface in essence
can still be thought of as intersecting 1D Fermi sheets, even after the
bands have been folded down into the lower symmetry cell (see text).\textit{
b)} A shift of one of the quasi-1D Fermi sheets (dashed lines represent the
shifted FS) along the (1,1,0) direction produces nearly perfect nesting. 
\textit{c)} A shift along the observed CDW wave vector direction (1,0,0)
produces imperfect nesting.}
\label{FS}
\end{figure}

Nearest neighbor Te ions are inequivalent due to symmetry breaking imposed
by the stacking of layers along the third direction. This results in a
larger unit cell and a smaller, rotated BZ, shown superimposed in Fig.\ref
{cete3struct}c. The full, calculated FS is shown in Fig.\ref{FS}. A very
good facsimile of it is achieved by folding the quasi-1D sheets of our TB
model down into the new zone. Though small gaps appear during the folding
down process, the long 1D ribbons are still clearly visible and both the
nesting peak and the peak in $\chi ^{\prime }(\mathbf{q})$ can be traced
back to these sheets.

As can be seen by the cartoons in Fig.\ref{FS}, there is an excellent (ideally perfect in the nearest-neighbor TB model) FS nesting 
between two of the quasi-1D ribbons along the (110) direction of the BZ. This produces very sharp nesting-derived peaks in $\chi^{\prime 
}(\mathbf{q})$, away from the observed nesting vector. A moderate peak at $\mathbf{q}$ = $\mathbf{q}_{CDW}$ is visible, but by far the 
strongest peaks are located elsewhere in the Brillouin zone. Any CDW directly stemming from FS nesting would occur first at wave vectors 
corresponding to these alternate spots rather than at the observed $\mathbf{q}_{CDW}$. In the real part of the susceptibility, the 
strongest FS nesting peaks are suppressed dramatically, leaving only a peak at the correct wave vector (Fig.\ref{FS}). This is entirely 
due to an effect of the finite-energy transitions (note the non-zero value of terms in Eq.  \ref{compsusc} even for widely spaced 
eigenvalues) that appear in the real part of susceptibility but not in the nesting function (imaginary part). The reason that the bulk of 
the $\chi^{\prime }(\mathbf{q})$ peak height at $ \mathbf{q}$ = $\mathbf{q}_{CDW}$ comes from contributions away from the Fermi energy 
can be understood by an examination of the band structure: the velocities of the electronic states connected by $\mathbf{q}_{CDW}$ ($ 
\epsilon_k$ and $\epsilon_{k+q}$) are nearly equal and opposite at the Fermi energy (note that for the wave vector $\mathbf{q}_{nest}$ in 
Fig.\ref{FS}b, the opposite is true, $v_k$ and $v_{k+q}$ have the same sign). Therefore nearby states $\epsilon_k+v\delta k$ and 
$\epsilon_{k+q}-v\delta k$ are also of equal energy and are connected by $\mathbf{q}_{CDW}$. These are located above and below the Fermi 
energy respectively and do not contribute to the nesting function but do contribute to the real part of the susceptibility. This 
phenomenon has been pointed out earlier in Ref. \onlinecite{Hidden} under the title of 'hidden nesting'. Obviously, if there are regions 
of the energy spectrum with equal and opposite velocities, they will contribute heavily to $\chi ^{\prime }(\mathbf{q})$ even if they are 
imperfectly nested at the Fermi energy itself. On the other hand, strong nesting between states exactly at the Fermi energy may die 
away quickly in other parts of the energy spectrum and contribute little to $\chi^{\prime }(\mathbf{q})$.  This is the essence of why 
well nested Fermi surfaces fail to produce a peak in $\chi 
^{\prime }(\mathbf{q})$ not only for CeTe$_3$ but also for the dichalcogenides discussed in the previous section. The Fermi surface is 
simply too small a part of the energy range from which $\chi ^{\prime }(\mathbf{q})$ collects to have a decisive effect.  

\begin{figure}[tbp]
\includegraphics[width = 0.95 \linewidth]{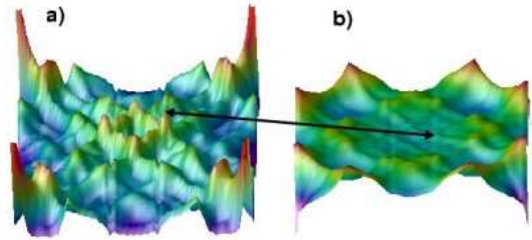}
\caption{(color online) A diagram showing the imaginary (top) and real
(bottom) parts of the susceptibility as a function of $q_x, q_y, q_z=0$. The
arrow connects the strongest peak in the imaginary part to its corresponding
position in the real part.}
\label{CeTe3}
\end{figure}

\subsection{Na chain}

\begin{figure}[t]
\includegraphics[width = 0.95\linewidth]{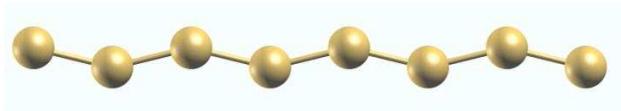}
\caption{The relaxed configuration of an initially one-dimensional chain of
Na ions. If the dimensionality of the chain is restricted, no distortion at
all occurs.}
\label{Nachain}
\end{figure}

To best approximate a perfect Peierls system, we calculated the ground state of a chain of Na ions separated by 38 $\AA$ in the two non-chain directions.  The Na-Na distance 
was relaxed along the chain to its optimal value of 3.34 $\AA$.  As we saw previously with NbSe$_2$, no CDW could be stabilized if the ions were clamped to their high 
symmetry positions, even though this system is ideal in every respect, \textit{i.e.} the electronic-only CDW is unstable even when no mitigating factors such as higher 
dimensionality or imperfect nesting are present.  Surprisingly, a relaxation of the ionic positions along the one-dimensional chain also failed to produce any deviation from 
the high symmetry, equally spaced arrangment, even when an initial dimerization was imposed. However, when the ions were allowed a larger dimensional freedom, the system 
distorted into a zigzag configuration as shown in Fig. \ref{Nachain}. The ions in the zigzag arrangement are 3.43 $\AA$ apart and the angle formed by the distortion is 152 
degrees.  Although this distortion doubles the unit cell and lowers the total energy, it does not create the expected gap at E$_F$.  The coupling between matrix elements at 
the BZ boundary (k=$\pm \pi/a$) that would give rise to a gapped system does not occur because the distortion is two dimensional, while the wavefunctions are 
one-diemnsional.  Thus, integration over the direction perpendicular to the chain in the dimension in which the distorion occurs drives the matrix element to zero.  
Precisely the same result can be obtained using a tight-binding formulation.  Regardless of the number of neighbors included in the model, the states at the edge of the BZ 
remain degenerate, {\it i.e.} there is no gap.  On the other hand, an enforced dimerization of the ions does produce a gap at E$_F$, but is unstable.   The true 
origin of the two-dimensional distortion is not yet entirely clear, but the Peierls mechanism can be ruled out. 
Thus, under perfect conditions, no electronic-only CDW forms, and no electron-ion interaction assisted CDW forms either. This indicates that the Peierls formulation is even 
weaker than we originally set out to prove. Not only do small deviations from the ideal system destroy the divergence that is purported to cause CDW formation, but in the 
ideal case itself, where no such deviations are present, the existence of ion cores is enough to effectively nullify the Peierls instability.

\section{Summary}

To summarize, we have first explored the standard half-filled
one-dimensional free electron system from the perspective of a Peierls
distortion and have shown that the divergence in the real part of the
susceptibility, caused by Fermi surface nesting, is exceedingly fragile.
Effects such as temperature, imperfect nesting, or scattering, all of which
are expected to be operative in a real material, reduce the divergence to a
simple peak, often not more than a factor of two or three enhanced over the
baseline susceptibility at \textbf{q}$\sim$0. Thus, expectations of a CDW
transition driven entirely by a Fermi surface nesting in any real material
are unrealistic from the outset. Next, we have examined a system in which a
gap at the Fermi surface has already been opened due a commensurate or
incommensurate lattice transition, and found that the energy gain comes
largely from the lowering of already filled states located \textit{away}
from the Fermi energy and not from removing states from the Fermi energy
itself. This reinforces our contention that the Fermi surface topology plays
at best a secondary role in CDW formation. Using first principles
calculations, we take three examples of well-known CDW materials to
illustrate our point and in each case we find that Fermi surface nesting
either does not exist at the CDW wave vector, or is not the strongest
nesting in the system. We further find that the CDW instability is not
fundamentally electronic, but rather stems from strong electron-phonon
interaction which, of course, is itself affected by the electronic
structure.. Finally, we examine the canonical Peierls system, a
one-dimensional chain of atoms, and find that the expected dimerization
along the chain axis with any realistic amplitude is energetically
unfavorable. The expected doubling of the unit cell occurs only if the
one-dimensionality is relaxed to allow a zigzag, rather than dimerized,
distortion. We conclude that no true distinction between CDWs and structural
phase transitions, in particular incommensurate lattice transitions, can be
made. Inspecting the Fermi surface itself for possible nesting features,
without actually calculating the real part of electronic susceptibility has
no predictive power for such structural transitions. Calculating the real
part of the electronic susceptibility may be helpful in analyzing such instabilities, but
only in relatively few cases can it be accepted as the only or even the main
driving force for such transitions.

\acknowledgments{We are especially grateful for the suggestions and advice of W.E. Pickett whose input has greatly improved this work.  We also acknowledge several useful 
and enjoyable discussions with I.R. Fisher and his collaborators at Stanford, as well as 
helpful communications with S.V. Borisenko.  Funding for 
research at NRL comes from the Office of Naval Research.}

\end{document}